\documentclass{article}
\usepackage{spconf,amsmath,graphicx,hyperref}

\usepackage{xcolor,soul,framed} 

\colorlet{shadecolor}{yellow}
\usepackage{array}
\usepackage{url}

\usepackage{cite}
\usepackage{booktabs}
\usepackage{balance} 
\usepackage{multirow}
\usepackage{multicol}
\usepackage{amsfonts}
\usepackage{float}
\usepackage{bbold}

\usepackage{hyperref}
\usepackage{etoolbox,siunitx}
\robustify\bfseries
\usepackage{balance}

\usepackage{amssymb}
\usepackage{pifont}
\usepackage{scalerel}

\title{Beyond Lips: Integrating Gesture and Lip Cues for Robust Audio-visual Speaker Extraction}
\name{Zexu Pan$^1$, Xinyuan Qian$^2$, Shengkui Zhao$^1$, Kun Zhou$^1$, Bin Ma$^1$}

\address{
  $^1$Tongyi Lab, Alibaba Group, Singapore \\
  $^2$University of Science and Technology Beijing, China
}

\begin{document}
\ninept
\maketitle
\setlength{\abovedisplayskip}{4pt}
\setlength{\belowdisplayskip}{4pt}

\begin{abstract}
Most audio-visual speaker extraction methods rely on synchronized lip recording to isolate the speech of a target speaker from a multi-talker mixture. However, in natural human communication, co-speech gestures are also temporally aligned with speech, often emphasizing specific words or syllables. These gestures provide complementary visual cues that can be especially valuable when facial or lip regions are occluded or distant. In this work, we move beyond lip-centric approaches and propose \textbf{SeLG}, a model that integrates both lip and upper-body gesture information for robust speaker extraction. SeLG features a cross-attention-based fusion mechanism that enables each visual modality to query and selectively attend to relevant speech features in the mixture. To improve the alignment of gesture representations with speech dynamics, SeLG also employs a contrastive InfoNCE loss that encourages gesture embeddings to align more closely with corresponding lip embeddings, which are more strongly correlated with speech. Experimental results on the YGD dataset, containing TED talks, demonstrate that the proposed contrastive learning strategy significantly improves gesture-based speaker extraction, and that our proposed SeLG model, by effectively fusing lip and gesture cues with an attention mechanism and InfoNCE loss, achieves superior performance compared to baselines, across both complete and partial (i.e., missing-modality) conditions.

\end{abstract}

\begin{figure*}[t]
\centering
\includegraphics[width=0.78\linewidth]{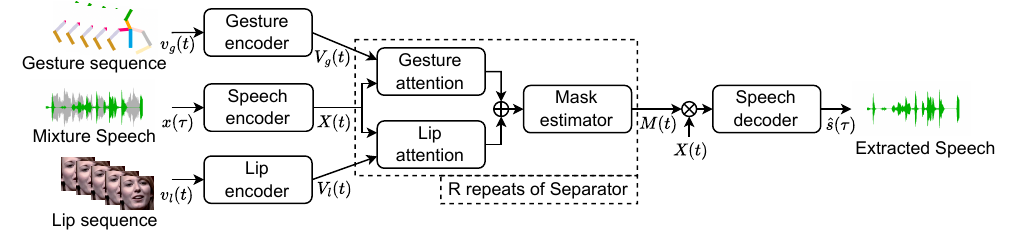}
\vspace{-3mm}
\caption{Our proposed audio-visual speaker extraction model named \textbf{SeLG}, which estimates the target speech conditioned on the synchronized gesture sequence $v_g(t)$ and the lip sequence $v_l(t)$ of the target speaker. The symbols $\oplus$ and $\otimes$ represent element-wise addition and multiplication, respectively.
}
\vspace{-3mm}
\label{fig:network}
\end{figure*}

\section{Introduction}
\label{sec:introduction}
Speech is the most natural form of human communication. However, speech signals are often contaminated by interfering background speakers, posing a significant challenge for speech processing tasks such as automatic speech recognition, speaker verification, or localization. Humans have the innate ability to focus attention on a speaker of interest, i.e., the target speaker, in multi-talker environments~\cite{bronkhorst2000cocktail,brungart2007cocktail}. Inspired by this capability, researchers have spent decades trying to mimic this behavior in machines, a challenge commonly referred to as the ``cocktail party problem''~\cite{ephrat2018looking}.

Speech separation is one prominent direction of research aimed at addressing the cocktail party problem. It seeks to decompose a multi-talker speech signal into individual clean streams by leveraging the unique dynamics or prosody of each speaker’s voice~\cite{hershey2016deep, luo2019conv, luo2020dual, wang2023tf, sajio2024locoformer}. However, the resulting separated streams lack explicit association with speaker identity, making it difficult to identify which output corresponds to the target speaker. Moreover, most methods require prior knowledge of the number of speakers for optimal performance, limiting their applicability in realistic, dynamic environments.

Speaker extraction algorithms, in contrast, more closely mimic the human ability to selectively extract the speech of a target speaker from a mixture, by leveraging an auxiliary cue to guide attention. Examples include pre-enrolled reference utterances~\cite{vzmolikova2019speakerbeam,wang2019voicefilter,spex_plus2020}, lip movements from video recordings~\cite{wu2019time,pan2021reentry,usev21}, directional cues~\cite{elminshawi2023beamformer}, subject's neural signals~\cite{biss2020,pan2023neuroheed,pan2024neuroheed+}, or other speaker-unique characteristics~\cite{tzinis2022heterogeneous,dai2025inter} to disentangle a specific voice from overlapping speech. Among these, visual cues are especially popular in human-robot interaction scenarios, as onboard cameras can easily capture synchronized video without requiring a pre-enrollment or selection process.

Among visual cues, the majority of existing work has focused on lip movements as the primary modality. However, the lips may often be occluded, or difficult to be detected and tracked reliably in low-resolution or distant-view scenarios. In contrast, the upper body remains more visible from a distance and is less prone to full occlusion. Embodied human communication involves a rich interplay between verbal (speech) and non-verbal behaviors, including body posture, hand gestures, and head nods~\cite{ahuja2020style,kendon2004gesture}. Notably, co-speech gestures are tightly synchronized with speech and highly correlated with its conceptual content and prosodic structure. Neuroscience studies further suggest that these gestures play a functional role in both speech production and perception, enhancing comprehension for listeners~\cite{wagner2014gesture,obermeier2015speaker}.

The SEG model has utilized co-speech gestures as a cue for target speaker extraction~\cite{pan2022seg}. However, the correlation between gesture and speech is less direct and weaker compared to the lip-speech relationship, leading to a performance gap. In this work, for the first time, we aim to leverage both lip and gesture cues for robust speaker extraction, which is particularly useful in scenarios where one modality is missing or degraded. 

The \textbf{SeLG} model (Speech extraction with Lip and co-speech Gesture cues) is proposed. We first introduce the cross-attention fusion method that uses each visual cue to query and select relevant speech features from the mixture as opposed to simple concatenation fusion in previous works~\cite{pan2022seg, usev21}. We also propose to enhance the discriminative representation of gesture embeddings to better correlate with the speech dynamics,  by aligning them with lip motion using an InfoNCE loss, which pulls gesture embeddings closer to corresponding lip embeddings. 

Experimental results on the ``in the wild'' YouTube Gesture Dataset (YGD)~\cite{yoon2019robots}, which includes TED talks, demonstrate that the proposed contrastive learning strategy significantly improves gesture-based extraction. Moreover, the attention-based fusion of lip and gesture cues outperforms the concatenation baseline, with the best performance achieved when combined with the InfoNCE loss. Improvements are observed in both two- and three-speaker mixtures, as well as across scenarios with and without missing modalities.

\section{Methodology}
Let $x(\tau)$\footnote{Variables indexed by $\tau$ denote time-domain signals, while those indexed by $t$ denote frame-based embeddings.} be a multi-talker mixture speech signal, composed of the target speech signal $s(\tau)$ and interference speech signal $b(\tau)$. The audio-visual speaker extraction (AVSE) network we studied in this work estimates the target speech signal $\hat{s}(\tau)$ to approximate $s(\tau)$, conditioned on the synchronized gesture sequence $v_g(t)$, or the combination of $v_g(t)$ and the lip sequence $v_l(t)$ of the target speaker.

\subsection{Model architecture}
Our proposed model is shown in Fig.~\ref{fig:network}, which consists of a gesture encoder, a lip encoder, a speech encoder, a speech decoder, and an ensemble separator. The separator comprises lip and gesture attention layers along with a mask estimator, as shown in the dotted box.

\subsubsection{Gesture encoder}
The input to the gesture encoder $v_g(t)$ is a sequence of human upper-body poses, represented as 3-dimensional (3D) coordinates of ten spine-centered joints: head, neck, nose, spine, left/right (L/R) shoulders, L/R elbows, and L/R wrists. The 3D coordinates are obtained by first applying a gesture estimation algorithm to the raw video to extract 2D pose sequences~\cite{yoon2019robots}, which are then lifted to 3D using a pose estimator~\cite{pavllo20193d}. The resulting 3D pose sequence $v_g(t)$ is passed through the gesture encoder, which is an $R_{g}$-layer Bidirectional Long Short-Term Memory (BLSTM), that models the temporal dynamics of gestures, yielding the encoded representation $V_g(t)$ that is in sync with the target's speech.

\subsubsection{Lip encoder}
The input to the lip encoder is a sequence of face images $v_l(t)$, extracted using a face detection and tracking network~\cite{liu2016ssd,Nagrani2017}. The sequence $v_l(t)$ is passed through the lip encoder to obtain a synchronized sequence of lip embeddings aligned with the target's speech. The lip encoder follows a structure similar to the visual encoder in MuSE~\cite{pan2020muse}, consisting of a 3D convolutional layer, followed by an 18-layer residual convolutional network pre-trained on a visual speech recognition task~\cite{stafylakis2017combining}, and then five stacked visual temporal convolutional network blocks~\cite{wu2019time}.

\subsubsection{Speech encoder and decoder}
The speech encoder consists of a 1D convolutional layer followed by a rectified linear unit activation, which converts the time-domain speech signal $x(\tau)$ into a spectrum-like, frame-based embedding sequence $X(t)$ in the latent space. The convolution uses an input channel size of 1, output channel size $N$, kernel size $L$, and stride $L/2$.

The speech decoder comprises a linear layer followed by an overlap-and-add operation to reconstruct the enhanced time-domain speech signal $\hat{s}(\tau)$. The input to the decoder is obtained by element-wise multiplication of the encoded features $X(t)$ and the estimated mask $\hat{M}(t)$, produced by $R$ repeated blocks of the separator (detailed in the next section). 

\vspace{-1mm}
\subsubsection{Separator}
\vspace{-1mm}
The separator is shown in the dotted box of Fig.~\ref{fig:network}, which aims to estimate the mask $\hat{M}(t)$ from the mixture embedding $X(t)$, guided by the gesture embedding $V_g(t)$ and lip embedding $V_l(t)$. We proposed to use a transformer cross-attention layer~\cite{rouard2022hybrid} to use each of the visual cues $V_l(t)$ or $V_g(t)$, to query $X(t)$ and select the correlated contents from $X(t)$ (where the query is from the visual embedding while key and values are from $X(t)$), and combine the attended embeddings from gesture cue and lip by element-wise addition. The resulting embedding is passed through a dual-path BLSTM module~\cite{luo2020dual,usev21}, to estimate the mask $M(t)$. The cross attention layers and the mask estimator are repeated $R$ times.

\subsection{Objective function}
Contrastive Predictive Coding (CPC)~\cite{oord2018representation,he2020momentum} is widely used in representation learning for extracting meaningful embeddings. It employs categorical cross-entropy to distinguish a positive sample from a set of noise (negative) samples, i.e.,  InfoNCE loss~\cite{oord2018representation,pan2023imaginenet}. In our design to improve gesture representation, the positive sample is the lip embedding at the corresponding time step, while the negative (i.e., non-matching) samples are lip embeddings from other time steps. The loss function is formulated as follows:
\begin{equation}
    \label{eqa:loss_infonce}
    \mathcal{L}_{emb} (V_l, V_g) = \sum_{i=1}^T -log \frac{exp(V_g^i \cdot V_l^i / \kappa)}{\sum_{j=1}^T exp(V_g^i \cdot V_l^j / \kappa)} 
\end{equation}
where $\kappa$ is the temperature parameter and set to 0.07, following the paper~\cite{he2020momentum}. The lip embedding $V_l$ is detached from the computational graph prior to computing this loss to prevent gradient backpropagation from altering the learned lip representation.

For enhancing the signal quality of the extracted signals, we use the negative scale-invariant signal-to-noise ratio (SI-SNR)~\cite{le2019sdr} as the objective function for training:
\begin{equation}
    \label{eqa:loss_SI-SNR}
    \mathcal{L}_{\text{SI-SNR}}(s, \hat{s}) = - 20 \log_{10} \frac{|\frac{<\hat{s},s>}{|s|^2}s|}{|\hat{s} - \frac{<\hat{s},s>}{|s|^2}s\big|}
\end{equation}

The total loss to train the models with additional contrastive loss is the summation of $\mathcal{L}_{emb}$ and $\mathcal{L}_{\text{SI-SNR}}$.

\section{Experimental setup}
\vspace{-1mm}
\subsection{Dataset}
\vspace{-1mm}
Following~\cite{pan2022seg}, we use the 2-speaker mixture (YGD-2mix) and the 3-speaker mixture (YGD-3mix) benchmark datasets, simulated using the source YouTube Gesture Dataset (YGD)~\cite{yoon2019robots,yoon2020speech}. The original dataset contains 27,611, 3,654, and 3,475 video segments in the train, validation, and test sets, respectively, drawn from 1,696 TED talks. In this work, we extend the dataset by extracting face recordings from the original videos. Approximately one-quarter of the videos have undetectable faces due to occlusion or distant camera views. For 20\% of the remaining videos where faces are visible, we randomly drop the gesture cues to simulate the missing modality scenarios, while ensuring that each video contains at least one visual modality, either lip or gesture.

All videos are sampled at 15 frames per second (FPS), resulting in synchronized 15 FPS sequences for both 3D poses and lip images. The audio is sampled at 16 kHz and converted to single-channel (mono) format. In total, 200,000, 5,000, and 3,000 mixture utterances are generated for the training, validation, and test sets, respectively, for both the YGD-2mix and YGD-3mix datasets. Each mixture is created by combining the target speech with one or more interfering speakers at a random Signal-to-Noise Ratio (SNR) uniformly sampled between $-10$ dB and $10$ dB. When the utterances have different durations, the longer speech is truncated to match the length of the shorter one. No speaker overlap exists across the train, validation, and test sets, enabling speaker-independent evaluation.

\vspace{-1mm}
\subsection{Baselines}
\vspace{-1mm}
We compare our method to the following baselines in this work:
\par
\noindent
\textbf{SEG}~\cite{pan2022seg}: An AVSE network that uses only gesture cues\footnote{The code is taken from~\url{https://github.com/modelscope/ClearerVoice-Studio}}. It employs concatenation-based fusion, where the gesture embedding $V_g(t)$ is upsampled and concatenated with the speech embedding $X(t)$ before being fed to the dual-path BLSTM mask estimator.
\par
\noindent
\textbf{USEV}~\cite{usev21}: An AVSE network that leverages only lip cues. Similarly, it adopts concatenation fusion by upsampling and concatenating the lip embedding $V_l(t)$ with $X(t)$ as input to the dual-path BLSTM mask estimator.
\par
\noindent
\textbf{SeLG$^{\circ}$ }: A variant of our proposed method that fuses lip and gesture cues via simple concatenation. Specifically, $V_l(t)$ and $V_g(t)$ are upsampled and concatenated with $X(t)$ and passed to the mask estimator. This baseline does not employ the InfoNCE loss for enhancing gesture representations.
\par
\noindent
\textbf{SeLG$^{\dagger}$ }: A variant of our proposed method that fuses lip and gesture cues via the cross-attention mechanism, but does not employ the InfoNCE loss for enhancing gesture representations.

\vspace{-1mm}
\subsection{Model configuration}
\vspace{-1mm}
For our SeLG network, the hyperparameters of the lip encoder are adopted from the USEV network~\cite{usev21}. For the gesture encoder, we use a bidirectional LSTM with $R_{g} = 5$ layers, a hidden size of 32, and a dropout of 0.3. The speech encoder and decoder are configured with
channel size $N = 256$ and a kernel stride of $L = 40$. The mask estimator has an input size of 64, a hidden size of 128, and a chunk size of 100 frames. For the cross-attention lip and gesture layer, it has an embedding dimension of 64, 4 attention heads, a feed-forward network dimension of 256, and a dropout of 0.3.

\subsection{Training}
For all model training, we use the AdamW optimizer with an initial learning rate of $5 \times 10^{-4}$. When the SeLG model includes a cross-attention layer, we apply a learning rate warm-up over the first 15,000 steps. The learning rate is halved if the best validation loss (BVL) does not improve for 6 consecutive epochs, and training is terminated if no improvement is observed for 10 epochs. Models are trained with an effective batch size of 64. To fit within GPU memory constraints, speech clips are truncated to a maximum duration of 10 seconds during training. For models trained with the additional $\mathcal{L}_{\text{emb}}$ loss, we initialize them from the corresponding checkpoint trained with only the $\mathcal{L}_{\text{SI-SNR}}$ loss and perform fine-tuning under the same training schedule.

\begin{table*}
    \centering
    \sisetup{
    detect-weight, 
    mode=text, 
    tight-spacing=true,
    round-mode=places,
    round-precision=1,
    table-format=2.2
    }
    \caption{Results on the 2-speaker mixture (YGD-2mix) and 3-speaker mixture (YGD-3mix) datasets. Each model is assigned a system number (Sys.\ \#), and performance is reported with respect to modality usage, fusion strategy, and loss function. The SI-SNRi (in dB) metric is first reported on the full test set, which includes both complete and missing-modality conditions. We further report SI-SNRi on two mutually exclusive subsets: (1) the \textit{w/o missing} subset, where all utilized visual cues are fully available; and (2) the \textit{w/ missing} subset, where either the lip or gesture cue is missing for multi-cue models (e.g., SeLG), and where the single utilized cue is missing for unimodal baselines.} 
    \begin{tabular}{cccccc SSSSSS}
       \toprule
        {Sys. \#}    &{Model}   &{Speakers}  &{Modalities} &{Fusion} &{Loss} &{Full test set}&{w/o missing}&{w/ missing}\\ 
        \midrule
         1      &USEV~\cite{usev21}         &\multirow{6}*{2}       &Lip        &Concatenation    &$\mathcal{L}_{\text{SI-SNR}}$     &8.5404     &13.0189    &-2.1969\\
         2      &SEG~\cite{pan2022seg}      &       &Gesture    &Concatenation    &$\mathcal{L}_{\text{SI-SNR}}$                     &6.976      &8.2708     &-1.3681\\
         3      &SEG      &       &Gesture    &Concatenation    &$\mathcal{L}_{\text{SI-SNR}}$ + $\mathcal{L}_{emb}$                 &7.6953     &8.9008     &-0.0732\\
         4      &SeLG$^{\circ}$             &       &Lip \& Gesture       &Concatenation    &$\mathcal{L}_{\text{SI-SNR}}$           &12.2723    &14.3682    &9.4789\\
         5      &SeLG$^{\dagger}$           &       &Lip \& Gesture       &Attention    &$\mathcal{L}_{\text{SI-SNR}}$               &12.8751    &14.985     &10.0630\\
         6      &SeLG   &   &Lip \& Gesture       &Attention    &$\mathcal{L}_{\text{SI-SNR}}$ + $\mathcal{L}_{emb}$                 &13.2081    &15.1102    &10.6729\\
        \midrule
         7      &USEV~\cite{usev21}         &\multirow{6}*{3}       &Lip        &Concatenation    &$\mathcal{L}_{\text{SI-SNR}}$     &9.5147     &13.0401    &0.2670\\
         8      &SEG~\cite{pan2022seg}      &       &Gesture    &Concatenation    &$\mathcal{L}_{\text{SI-SNR}}$                     &4.039      &4.5552     &0.7963\\
         9      &SEG      &       &Gesture    &Concatenation    &$\mathcal{L}_{\text{SI-SNR}}$ + $\mathcal{L}_{emb}$                 &4.7037     &5.3505     &0.6408\\
         10     &SeLG$^{\circ}$         &       &Lip \& Gesture       &Concatenation    &$\mathcal{L}_{\text{SI-SNR}}$               &11.1365    &14.82      &5.9085\\
         11     &SeLG$^{\dagger}$        &       &Lip \& Gesture       &Attention    &$\mathcal{L}_{\text{SI-SNR}}$                  &11.3467    &14.82      &6.3404\\
         12     &SeLG      &       &Lip \& Gesture       &Attention    &$\mathcal{L}_{\text{SI-SNR}}$ + $\mathcal{L}_{emb}$          &11.6687    &14.9028    &7.0784\\
        \bottomrule
    \end{tabular}
    \vspace{-3mm}
    \label{tab:results}
\end{table*}

\vspace{-1mm}
\section{Results}
\vspace{-1mm}
We compare our proposed SeLG model with the baselines using the improvement in SI-SNR (SI-SNRi), measured on the extracted speech signal with respect to the unprocessed signal. All models are trained on the same simulated YGD mixture dataset, in which certain samples have missing visual cues, either gesture or lip.  We evaluate performance on the full test set and further decompose the results into two mutually exclusive subsets: (1) the \textit{w/o missing} subset, where all utilized visual cues are fully available; and (2) the \textit{w/ missing} subset, where either the lip or gesture cue is missing for multi-cue models (e.g., SeLG), and the single employed cue is absent for unimodal baselines. This evaluation protocol enables a detailed analysis of model robustness under partial observability.

\vspace{-1mm}
\subsection{Effects of using both lip and gesture cues}
In Table~\ref{tab:results}, we first present all model results on the 2-speaker dataset YGD-2mix. For System 1 (lip-only), the SI-SNRi reaches 13.0~dB when lip cues are fully available (w/o missing), but drops to 8.5~dB on the full test set, which includes samples with missing lip cues. Notably, when the lip cue is fully absent, performance degrades significantly, with an SI-SNRi of only $-2.2$~dB. Similarly, System 2 (gesture-only) achieves 8.3~dB in the complete-modality setting, but performance decreases to 7.0~dB on the full test set, and drops to $-1.4$~dB when the gesture cue is fully missing. In contrast, System 4, which fuses both lip and gesture cues by concatenation, achieves 12.3~dB on the full test set, and 14.4~dB on the subset with both cues present. Even in missing-modality conditions, it maintains an average SI-SNRi of 9.5~dB, significantly outperforming both unimodal baselines. This demonstrates that integrating complementary visual modalities enhances robustness and accuracy in speaker extraction, particularly when one modality is unavailable.
A similar trend is observed on the 3-speaker dataset YGD-3mix, comparing System 7 (lip-only), System 8 (gesture-only), and System 10 which combines both visual cues, which system 10 shows robustness to its unimodal counterparts.

\vspace{-1mm}
\subsection{Effects of the attention fusion}

In Table~\ref{tab:results}, our proposed System 5, which uses attention-based fusion instead of concatenation (System 4), achieves a 0.6~dB improvement in SI-SNRi on the full test set for the 2-speaker scenario. This gain is consistent across both subsets with or without missing modalities, demonstrating the effectiveness of the attention mechanism in leveraging visual cues under all conditions.

For the 3-speaker scenario, System 11 (attention fusion) achieves a 0.2~dB improvement in SI-SNRi over its concatenation-based counterpart (System 10) on the full test set. A closer analysis reveals that this improvement stems primarily from the \textit{w/ missing} subset, while performance on the \textit{w/o missing} subset remains nearly unchanged. This indicates that our attention-based fusion mechanism is particularly effective in handling missing-modality conditions, where it better exploits available cues to maintain robust extraction performance.

\begin{figure}[t]
\begin{minipage}[t]{.48\linewidth}
  \centering
  \centerline{\includegraphics[width=\linewidth]{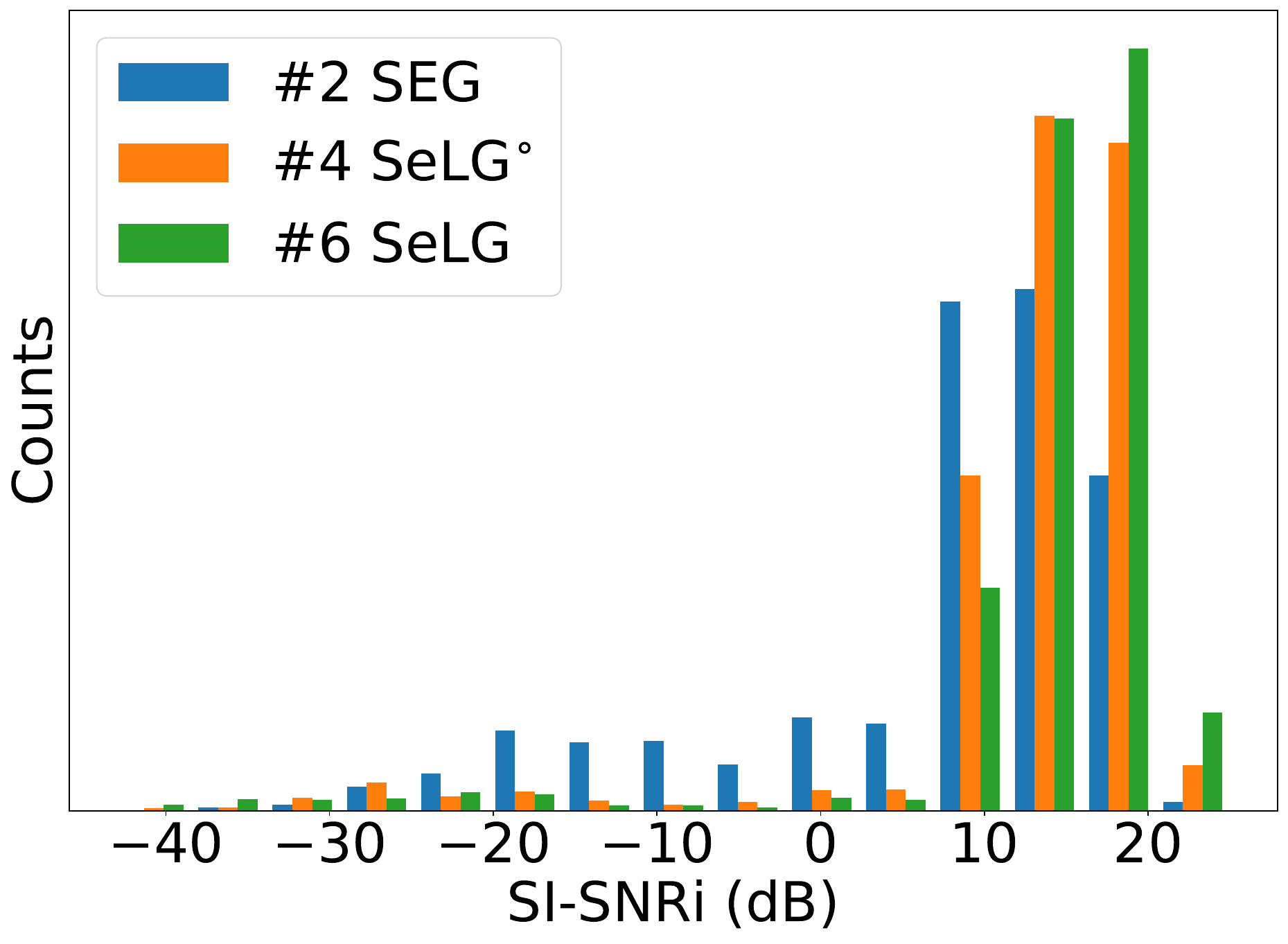}}
  \vspace*{-2mm}
  \caption{The SI-SNRi histogram of YGD-2mix test samples.}\medskip
  \label{fig:histogram_2spk}
\end{minipage}
\hfill
\begin{minipage}[t]{.48\linewidth}
  \centering
  \centerline{\includegraphics[width=\linewidth]{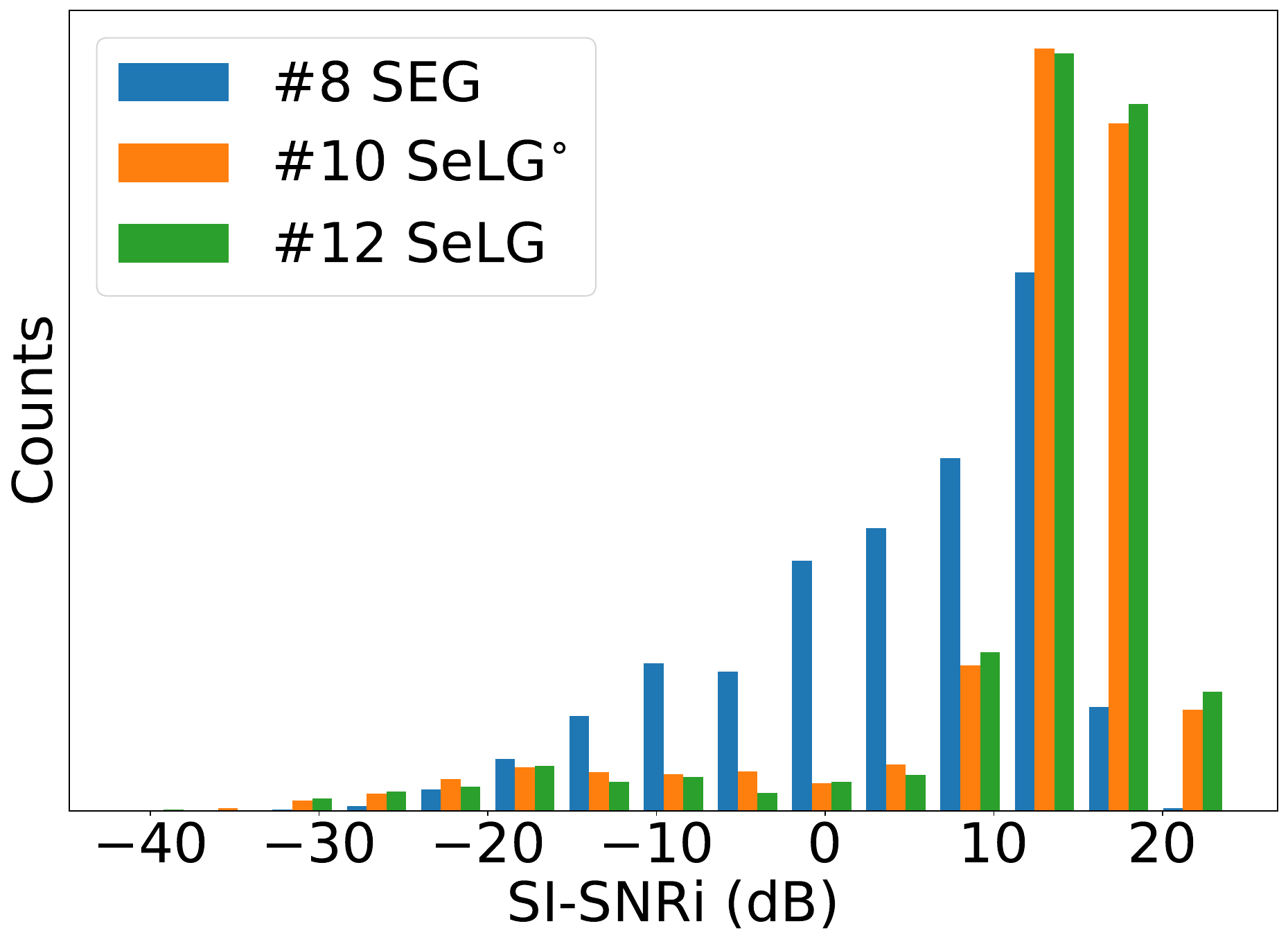}}
  \vspace*{-2mm}
  \caption{The SI-SNRi histogram of YGD-3mix test samples.}\medskip
  \label{fig:histogram_3spk}
\end{minipage}
\vspace*{-5mm}
\end{figure}

\subsection{Effects of using contrastive loss}

We also investigate the effectiveness of the proposed InfoNCE loss in improving gesture representations. We first study the SEG networks which only utilize the gesture cue for speaker extraction, the visual embedding used to compute InfoNCE loss is from the corresponding lip-only USEV models. For the 2-speaker scenario, comparing System 3 (with InfoNCE loss) to System 2 (without), we observe a 0.7~dB improvement in SI-SNRi, demonstrating that contrastive learning enhances the discriminative power of gesture features. A similar trend is observed in the 3-speaker scenario: System 9 (with InfoNCE) achieves a 0.7~dB gain over System 8 (without), further confirming the effectiveness of the InfoNCE loss in learning robust and speech-aligned gesture representations across different mixture complexities. 

For SeLG networks that utilize both lip and gesture cues, we first evaluate the impact of the proposed InfoNCE loss in the two-speaker scenario. System 6 (with InfoNCE) outperforms System 5 (without) by 0.3~dB on the full test set. A breakdown of the results shows only a 0.1~dB improvement on the \textit{w/o missing} subset, while a more substantial gain of 0.6~dB is observed on the \textit{w/ missing} subset. This indicates that the InfoNCE loss is particularly beneficial in missing-modality conditions, where it enhances the alignment and reliability of gesture representations when one cue is absent.
A similar trend is observed in the 3-speaker scenario: System 12 (with InfoNCE) outperforms System 11 (without), with performance gains concentrated in the \textit{w/ missing} subset. The consistent improvement under partial observability further demonstrates the effectiveness of the InfoNCE loss in learning robust and speech-aligned multimodal representations, especially when critical visual inputs are unavailable.

\vspace{-1mm}
\subsection{Sample distribution}

We also present the histogram of SI-SNRi across all test samples for the 2-speaker scenario in Fig.~\ref{fig:histogram_2spk} and the 3-speaker scenario in Fig.~\ref{fig:histogram_3spk}. It is observed that the proposed SeLG model exhibits a similar overall distribution to the SeLG$^{\circ}$ model, but the SeLG model has a greater concentration of samples at higher SI-SNRi values and fewer samples at lower SI-SNRi values in both scenarios. In contrast, the gesture-only (SEG) model shows a significant number of samples clustered between $-20$~dB and $-10$~dB, indicating frequent failure cases where the model extracts the wrong speaker, particularly when gesture cues are missing or exhibit weaker correlation with speech in the absence of lip information, leading to incorrect speaker selection during fusion.

\vspace{-1mm}
\section{Conclusion}

In this work, we proposed \textbf{SeLG}, a robust audio-visual speaker extraction framework that goes beyond lip-centric approaches by integrating both lip and upper-body co-speech gestures. We introduced a contrastive InfoNCE loss to align gesture embeddings with lip dynamics, leveraging the stronger correlation between lip movements and speech, thereby enhancing the discriminative power of gesture representations. Additionally, we designed a cross-attention-based fusion mechanism that effectively combines both visual modalities to extract the target speaker’s speech. Experiments on the YGD-2mix and YGD-3mix datasets demonstrate that our method significantly improves gesture-based extraction and outperforms unimodal and concatenation-based baselines, achieving superior performance even when one modality is missing. These results highlight the value of co-speech gestures as complementary cues and establish SeLG as a robust solution for speaker extraction under partial observability.

\newpage

\footnotesize
\bibliographystyle{IEEEbib}
\bibliography{IEEEabrv,Bibliography}

\end{document}